\documentclass{aa}

\usepackage[varg]{txfonts}
\usepackage[colorlinks=true, linkcolor=gray, citecolor=gray, urlcolor=gray]{hyperref}
\usepackage{floatrow}
\usepackage{xcolor}

\begin{document}

\title{Nonparametric galaxy morphology from UV to submm wavelengths}

\author{%
Maarten~Baes\inst{\ref{UGent}}
\and
Angelos~Nersesian\inst{\ref{NOA}}
\and
Viviana~Casasola\inst{\ref{INAFB}}
\and
Simone~Bianchi\inst{\ref{Firenze}}
\and
Letizia~P.~Cassar\`a\inst{\ref{NOA},\ref{Milano}}
\and
Christopher~J.~R.~Clark\inst{\ref{STScI}}
\and
Ilse~De~Looze\inst{\ref{UGent},\ref{UCL}}
\and
Wouter~Dobbels\inst{\ref{UGent}}
\and
Jacopo~Fritz\inst{\ref{Mex}}
\and
Maud~Galametz\inst{\ref{CEA}}
\and
Fr\'ed\'eric~Galliano\inst{\ref{CEA}}
\and
Suzanne~C.~Madden\inst{\ref{CEA}}
\and
Aleksandr~V.~Mosenkov\inst{\ref{RAS},\ref{SPB}}
\and
S\'ebastien~Viaene\inst{\ref{UGent},\ref{Heref}}
\and
Ana~Tr\v{c}ka\inst{\ref{UGent}}
\and
Emmanuel~M.~Xilouris\inst{\ref{NOA}}
}

\institute{
Sterrenkundig Observatorium, Universiteit Gent, Krijgslaan 281 S9, B-9000 Gent, Belgium
\label{UGent}
\and
National Observatory of Athens, Institute for Astronomy, Astrophysics, Space Applications and Remote Sensing, Ioannou Metaxa and Vasileos Pavlou GR-15236, Athens, Greece
\label{NOA}
\and
INAF -- Istituto di Radioastronomia, Via P. Gobetti 101, 40129, Bologna, Italy
\label{INAFB}
\and
INAF -- Osservatorio Astrofisico di Arcetri, Largo E. Fermi 5, 50125 Firenze, Italy
\label{Firenze}
\and
INAF -- Istituto di Astrofisica Spaziale e Fisica cosmica, Via A. Corti 12, 20133, Milano, Italy
\label{Milano}
\and
Space Telescope Science Institute, 3700 San Martin Drive, Baltimore, MD 21218, USA
\label{STScI}
\and
Department of Physics and Astronomy, University College London, Gower Street, London WC1E 6BT, UK
\label{UCL}
\and
Instituto de Radioastronom\'{\i}a y Astrof\'{\i}sica, UNAM,  
Antigua Carretera a P\'{a}tzcuaro \# 8701, Ex-Hda. San Jos\'{e} de la Huerta, 58089 Morelia, Michoac\'{a}n,  Mexico 
\label{Mex}
\and
AIM, CEA, CNRS, Universit\'e Paris-Saclay, Universit\'e Paris Diderot, Sorbonne Paris Cit\'e, 91191 Gif-sur-Yvette, France
\label{CEA}
\and
Central Astronomical Observatory of RAS, Pulkovskoye Chaussee 65/1, 196140 St. Petersburg, Russia
\label{RAS}
\and
St. Petersburg State University, Universitetskij Pr. 28, 198504 St. Petersburg, Stary Peterhof, Russia
\label{SPB}
\and
Centre for Astrophysics Research, University of Hertfordshire, College Lane, Hatfield, AL10 9AB, UK
\label{Heref}
}

\date{Received 22 May 2020 / Accepted 3 July 2020}

\abstract{We present the first nonparametric morphological analysis of a set of spiral galaxies from UV to submillimeter (submm) wavelengths. Our study is based on high-quality multi-wavelength imaging for nine well-resolved spiral galaxies from the DustPedia database, combined with nonparametric morphology indicators calculated in a consistent way using the {\tt{StatMorph}} package. We measure the half-light radius, the concentration index, the asymmetry index, the smoothness index, the Gini coefficient, and the $M_{20}$ indicator in various wavebands from UV to submm wavelengths, and in stellar mass, dust mass, and star formation rate maps. We find that the interstellar dust in galaxies is distributed in a more extended, less centrally concentrated, more asymmetric, and more clumpy way than the stars are. This is particularly evident when comparing morphological indicators based on the stellar mass and dust mass maps. This should serve as a warning sign against treating the dust in galaxies as a simple smooth component. We argue that the nonparametric galaxy morphology of galaxies from UV to submm wavelengths is an interesting test for cosmological hydrodynamics simulations.}


\maketitle

\section{Introduction}

The morphology of galaxies is essential to understand how galaxies form and evolve. Galaxy morphology is the basis of the standard classification schemes \citep{2005ARA&A..43..581S} and correlates with a wide range of physical properties, such as optical color, stellar mass, star formation history, and local environment  \citep{1980ApJ...236..351D, 2003ApJS..147....1C, 2005ApJ...625..621B, 2009MNRAS.399..966S, 2014MNRAS.441..599B}. In the current era of massive large-area surveys, an objective and automated determination of galaxy morphology is required. 

\begin{table*}
\caption{Basic properties of the galaxies in our sample. Redshift-independent distances $D$ and RC3 types are taken from the NED database; Hubble stages $T$ are from \citet{1988ngc..book.....T};  $R_{25}$ radii are taken from HyperLeda; and inclinations are compiled from different sources, as listed in the caption of Table~1 of \citet{2017A&A...605A..18C}. Stellar masses, dust masses, and SFRs are based on the CIGALE SED fits presented by \citet{2019A&A...624A..80N}. The pixel scale corresponds to the common 12\arcsec\ resolution used in our analysis.}
\label{Galaxies.tab}
\centering
\begin{tabular}{cccccccccccc}
\hline\hline \\[-0.5ex]
NGC & M & $D$ & $T$ & RC3 type & $R_{25}$ & $i$ & $\log M_\star$ & $\log M_{\text{dust}}$ & $\log {\text{SFR}}$ & pixel scale \\
& & [Mpc] &  &  & [arcmin] & [deg] & [$M_\odot$] & [$M_\odot$] & [$M_\odot$\,yr$^{-1}$] & [kpc/pix]
\\[1.5ex] \hline \\[-0.5ex]
NGC\,0628 & M\,74 & 9.0 & 5 & SA(s)c & 5.0 & 7 & 10.15 & 7.58 & 0.382 & 0.52\\
NGC\,2403 & -- & 3.5 & 6 & SAB(s)cd & 10.0 & 63 & 9.47 & 6.64 & --0.135 & 0.20\\
NGC\,3031 & M\,81 & 3.7 & 2 & SA(s)ab & 10.7 & 59 & 10.65 & 7.02 & --0.462 & 0.22 \\
NGC\,3521 & -- & 12.0 & 4 & SAB(rs)bc & 4.2 & 73 & 10.93 & 7.73 & 0.499 & 0.70 \\
NGC\,3621 & -- & 6.9 & 7 & SA(s)d & 4.9 & 65 & 10.31 & 7.03 & 0.153 & 0.40 \\
NGC\,4725 & -- & 13.6 & 2 & SAB(r)ab pec & 4.9 & 54 & 10.87 & 7.39 & --0.033 & 0.79 \\
NGC\,4736 & M\,94 & 5.2 & 2 & (R)SA(r)ab & 3.9 & 41 & 10.39 & 6.39 & --0.261 & 0.30 \\
NGC\,5055 & M\,63 & 8.2 & 4 & SA(s)b & 5.9 & 59 & 10.77 & 7.65 & 0.391 & 0.48 \\
NGC\,5457 & M\,101 & 7.0 & 6 & SAB(rs)cd & 12.0 & 18 & 10.15 & 7.67 & 0.677 & 0.41 \\[1.5ex] \hline\hline
\end{tabular}
\end{table*}

While novel approaches are being proposed on a regular basis \citep[e.g.,][]{2015ApJS..221....8H, 2015MNRAS.450.1441D, 2018MNRAS.476.5516B, 2020MNRAS.491.1408M, 2020MNRAS.493.4209C, 2020arXiv200406734U}, the most popular way to quantify galaxy morphology is to make use of morphological features or indicators. These indicators are often the parameters of analytical models fitted to the surface brightness distribution of galaxies, such as effective radii and S\'ersic parameters derived from single-component S\'ersic model fits \citep{2012MNRAS.427.1666B, 2012ApJS..203...24V, 2015MNRAS.454..806K} or bulge-to-disk  ratios from multi-component image decompositions \citep{2003ApJ...582..689M, 2009MNRAS.393.1531G, 2016MNRAS.462.1470L, 2017A&A...598A..32M, 2018ApJ...862..100G, 2018MNRAS.473.4731K, 2019MNRAS.486..390B}. 

An alternative method consists of the use of nonparametric morphological indicators, which are indicators that do not assume a fixed functional form of the surface brightness distribution. The most popular sets of nonparametric indicators are the concentration--asymmetry--smoothness (CAS) indices \citep{2000AJ....120.2835A, 2000AJ....119.2645B, 2003ApJS..147....1C} and the Gini--$M_{20}$ indices \citep{2003ApJ...588..218A, 2004AJ....128..163L}. Both sets have been widely applied to optical and near-IR (NIR) images of large samples of galaxies \citep{2006ApJ...636..592L, 2007ApJS..172..406S, 2009MNRAS.394.1956C, 2009A&A...497..743H, 2014ApJ...781...12H, 2014ARA&A..52..291C, 2019MNRAS.483.4140R}. One interesting aspect of nonparametric morphological indicators is that they can easily be applied to any image, and not only those dominated by stellar emission. As such, they are ideal tools to objectively and quantitatively compare the morphology of different galaxy components \citep{2007MNRAS.380.1313B, 2009ApJ...703.1569M, 2011MNRAS.416.2401H, 2016A&A...591A...1P}. 

No statistical nonparametric morphological studies have been made that extend into the far-infrared (FIR) and submillimeter (submm) regime. \citet{2009ApJ...703.1569M} do present measurements in the {\it{Spitzer}} MIPS 70 and 160 $\mu$m bands, but indicate that most of these values should be taken with caution. Thanks to the {\it{Herschel}} mission, these studies can be extended to the submm range where the emission of cold dust dominates. The angular resolution and sensitivity of {\it{Herschel}} are sufficient to spatially resolve large nearby galaxies, and several detailed studies of the cold dust emission within galaxies have been presented \citep[e.g.,][]{2010A&A...518L..69B, 2010A&A...518L..65B, 2012MNRAS.419.1833B, 2012ApJ...756..138A, 2020ApJ...889..150A, 2012MNRAS.421.2917F, 2012ApJ...755..165M, 2012ApJ...756...40S, 2014A&A...565A...4H, 2018A&A...616A.120M, 2019A&A...622A.132M}. 

In this paper we use multi-wavelength imaging from the DustPedia database\footnote{\url{http://dustpedia.astro.noa.gr}} \citep{2017PASP..129d4102D, 2018A&A...609A..37C} to present, for the first time, a nonparameteric morphological analysis of spiral galaxies from UV to submm wavelengths. Our goal is to consistently measure how galaxy morphology changes as a function of wavelength over this wide wavelength range. We also apply the same analysis to stellar mass, dust mass, and star formation rate (SFR) maps to directly quantify the morphology of different physical galaxy components. In Sect.~{\ref{Data.sec}} we discuss the sample and the data, in Sect.~{\ref{Methodology.sec}} we present the methodology we use for our analysis, in Sect.~{\ref{Results.sec}} we present our results, and in Section~{\ref{Discussion.sec}} we discuss the implications of our study.

\section{Sample and data}
\label{Data.sec}

For our analysis, we require an imaging data set of galaxies that satisfies the following conditions: (1)~the galaxies should have high-quality imaging data available from UV to submm wavelengths, (2)~the galaxies should be large enough on the sky to guarantee enough angular resolution out to 500~$\mu$m, and (3) the galaxies should not be too inclined to avoid strong projection effects. 

The DustPedia project \citep{2017PASP..129d4102D} was designed to gather a coherent multi-wavelength imaging data set for all nearby galaxies observed with {\it{Herschel}} and to analyze these data consistently with state-of-the-art modeling tools.  Some salient results of the DustPedia project include an analysis of the dust heating mechanisms in galaxies \citep{2019A&A...624A..80N, 2020A&A...637A..25N, 2020A&A...637A..24V, 2020arXiv200501720V}, determinations of dust absorption cross section within and between galaxies \citep{2019A&A...631A.102B, 2019MNRAS.489.5256C}, and an investigation of the ISM scaling relations in different environments \citep{2019A&A...626A..63D, 2020A&A...633A.100C}.
 
\begin{figure*}
\centering
\includegraphics[width=0.85\textwidth]{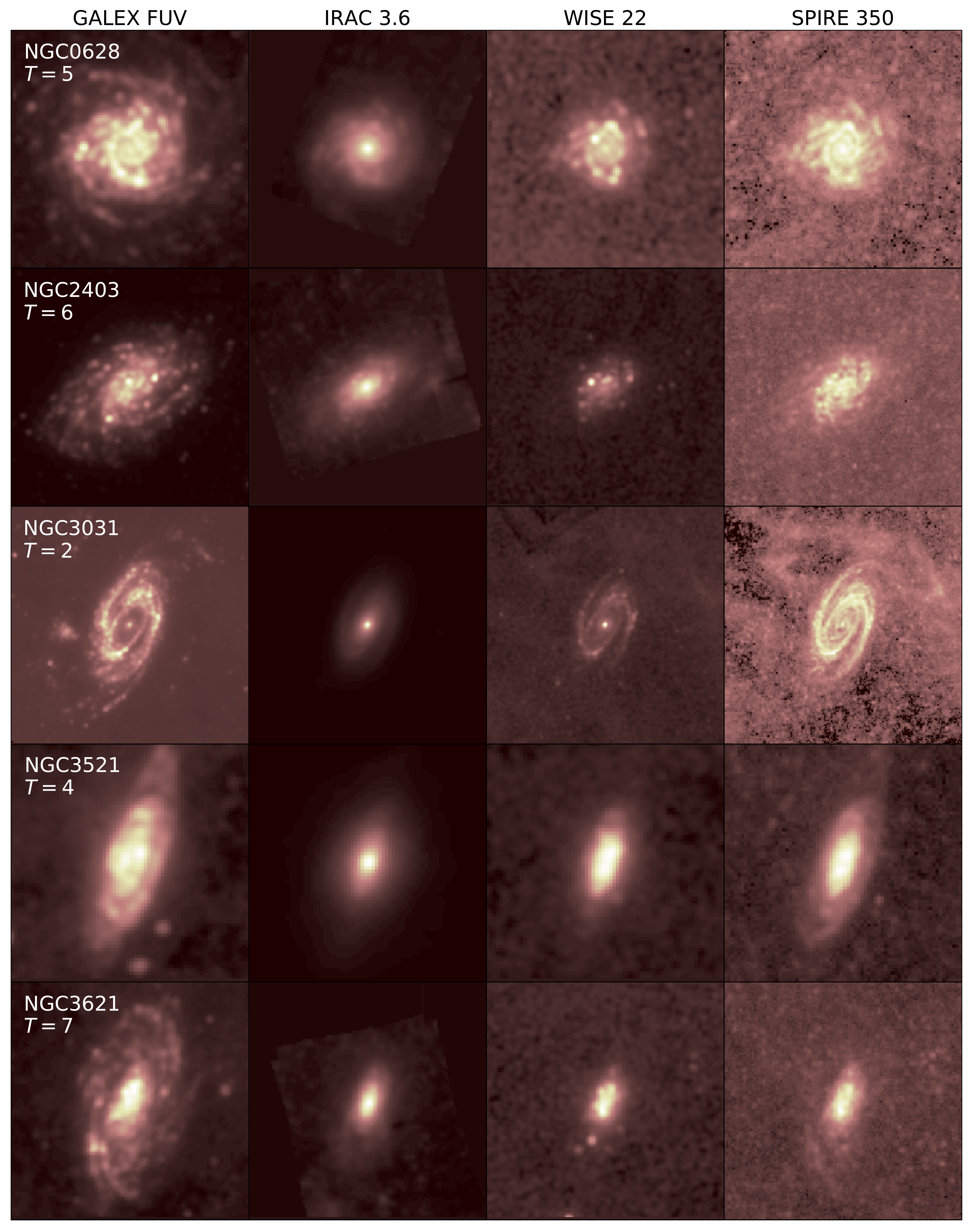}
\caption{{\em{GALEX}} FUV, IRAC 3.6 $\mu$m, {\em{WISE}} 22~$\mu$m, and SPIRE 350~$\mu$m images for NGC\,0628 (M\,74), NGC\,2403, NGC\,3031 (M\,81), NGC\,3521, and NGC\,3621. All images are convolved to the same FWHM and are shown on a square root scale.}  
\label{images1.fig}
\end{figure*}

\begin{figure*}
\centering
\includegraphics[width=0.85\textwidth]{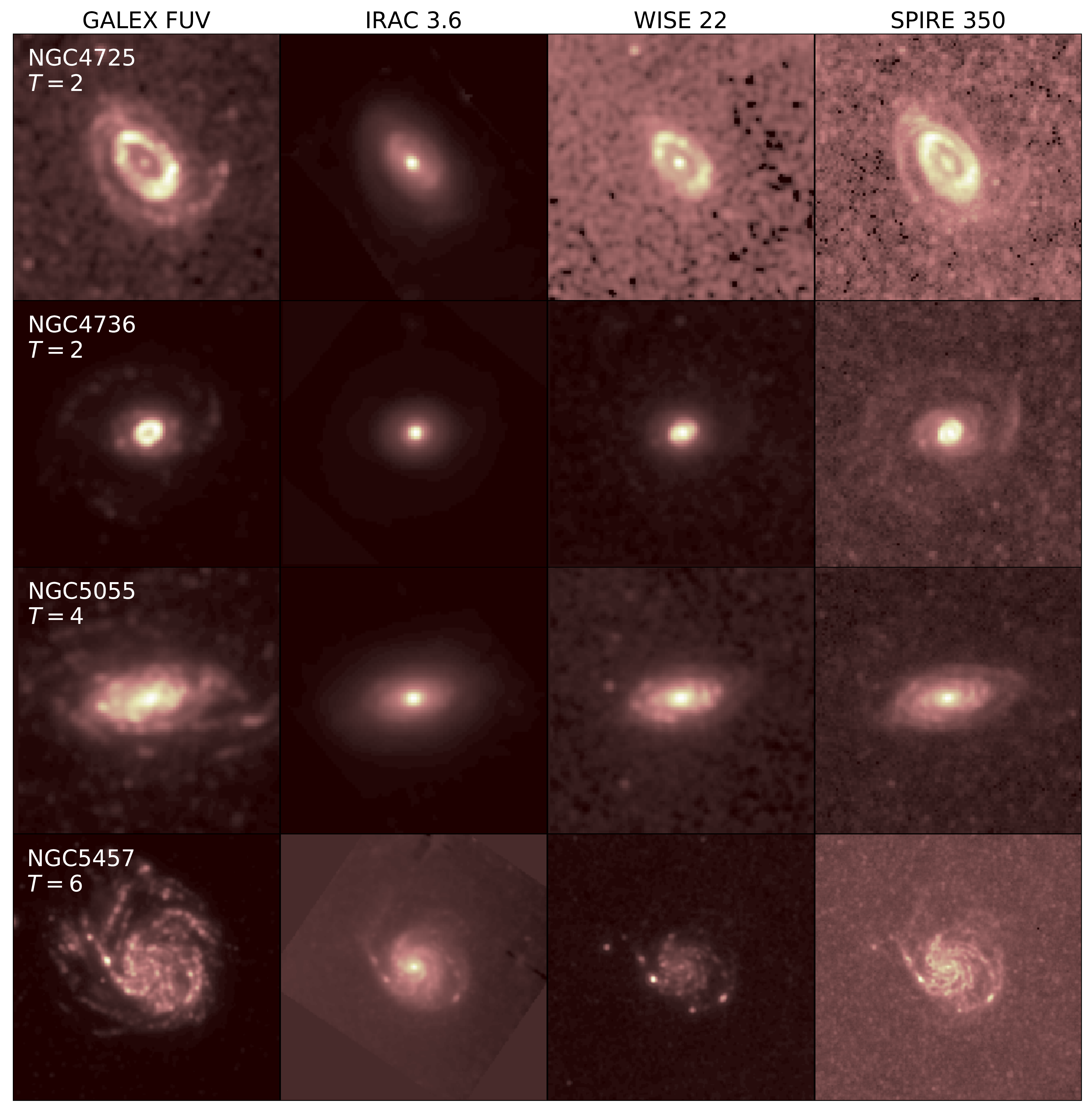}
\caption{{\em{GALEX}} FUV, IRAC 3.6 $\mu$m, {\em{WISE}} 22~$\mu$m, and SPIRE 350~$\mu$m images for NGC\,4725, NGC\,4736 (M\,94), NGC\,5055 (M\,63), and NGC\,5457 (M\,101). All images are convolved to the same FWHM and are shown on a square root scale.}  
\label{images2.fig}
\end{figure*}

The current project is based on the large face-on galaxy subsample considered by \citet{2017A&A...605A..18C}. They selected 18 galaxies from the DustPedia database with {\it{Herschel}} PACS and SPIRE imaging and with a submm diameter larger than 9~arcmin and axis ratio larger than 0.4. From this sample we selected the nine galaxies for which consistent imaging data were available over the entire UV--submm wavelength range with a smooth and sufficiently large background area in each band. Details on the galaxies considered are provided in Table~{\ref{Galaxies.tab}}. This sample, while small, contains galaxies with Hubble types ranging from $T=2$ to $T = 7$, and covers 1.5 orders of magnitude in stellar mass, and more than an order of magnitude in dust mass and SFR. 

For all galaxies in our sample, we use the same imaging data as used by \citet{2017A&A...605A..18C}. We use images in the {\it{Galaxy Evolution Explorer (GALEX)}} far-UV (FUV) and near-UV (NUV) bands, the Sloan Digital Sky Survey (SDSS) {\it{g}} and {\it{i}} bands, the {\it{Spitzer}} IRAC 3.6 and 4.5 $\mu$m bands, the {\it{Wide-field Infrared Survey Explorer (WISE)}} 12 and 22 $\mu$m bands, the three {\it{Herschel}} PACS bands and the three {\it{Herschel}} SPIRE bands. For some galaxies, imaging data in the SDSS bands was not available, in which case we used the available $B$ and $R$ band images from the Spitzer Infrared Nearby Galaxy Survey (SINGS) \citep{2003PASP..115..928K} instead.

Foreground stars were removed from the {\it{GALEX}}, SDSS, and IRAC images based on the 2MASS All-Sky Catalog of Point Sources \citep{2003tmc..book.....C}. All images were background subtracted and corrected for Milky Way attenuation. For more details on the data reduction and processing, we refer to \citet{2017A&A...605A..18C} and \citet{2018A&A...609A..37C}.

The final step in our image processing consists of convolving all images to the SPIRE 500~$\mu$m PSF using the convolution kernels of \citet{2011PASP..123.1218A} and re-gridding them on the same 12\arcsec\ pixel scale. The goal of our paper is to investigate how the morphology of galaxies changes as a function of wavelength, at a fixed resolution. By convolving all images to the same PSF, we separate the intrinsic wavelength-dependent effects from potential additional effects due differences in angular resolution. Systematic effects on nonparametric morphological indicators due to changes in angular resolution have been studied in detail by various authors \citep[e.g.,][]{2004AJ....128..163L, 2007MNRAS.380.1313B, 2011MNRAS.416.2401H}. These studies generally agree that the morphological parameters are not affected dramatically, as long as features smaller than about 1 kpc are resolved in the images. Our galaxies satisfy this criterion (see Table~{\ref{Galaxies.tab}}).

Apart from the monochromatic broadband images, we also considered stellar mass, dust mass, and SFR maps, which  were generated as described in \citet{2017A&A...605A..18C}. In short, the stellar mass maps were created from the IRAC 3.6 and 4.5 $\mu$m images following the prescriptions from \citet{2015ApJS..219....5Q}. The method used to create these maps is based on the Independent Component Analysis method and disentangles the contribution of old stellar populations, PAH emission, and hot dust emission \citep{2012ApJ...744...17M}. The dust mass maps were created by fitting THEMIS model dust SEDs \citep{2017A&A...602A..46J} to each set of corresponding pixels in the PACS and SPIRE maps. Finally, SFR surface density maps were created combining {\it{GALEX}} FUV and {\it{WISE}} 22~$\mu$m maps according to the prescriptions of \citet{2008AJ....136.2846B} and \citet{2008AJ....136.2782L}.

Figures~{\ref{images1.fig}} and {\ref{images2.fig}} show an overview of the  {\em{GALEX}} FUV, IRAC 3.6 $\mu$m, {\em{WISE}} 22~$\mu$m, and SPIRE 350~$\mu$m images for all the galaxies in our sample. All images are convolved to the same FWHM and shown on a square root scale.

\section{Methodology}
\label{Methodology.sec}

\begin{figure*}
\centering
\includegraphics[width=\textwidth]{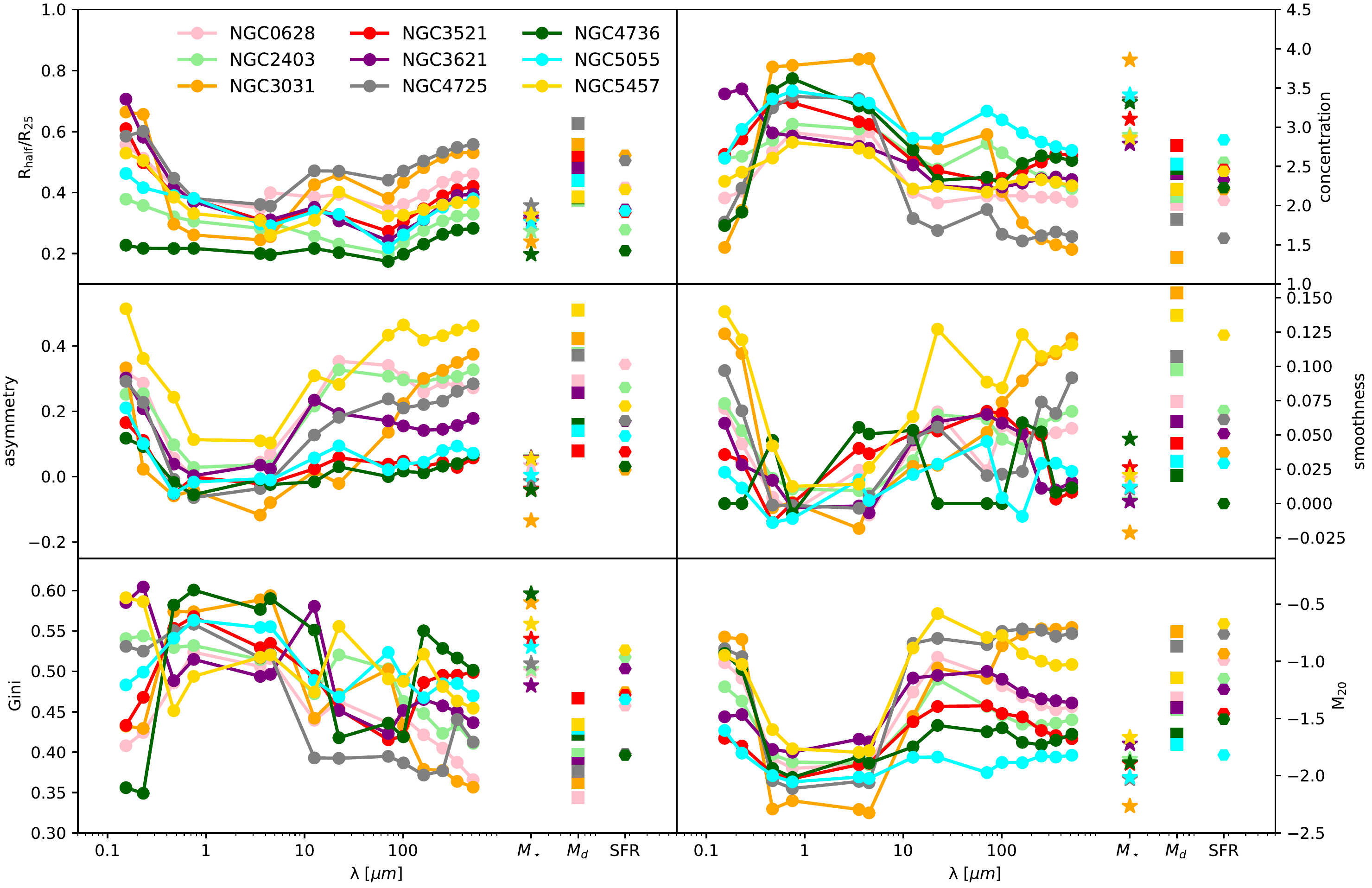}
\caption{Nonparametric morphological indicators for each of the galaxies in our sample as a function of wavelength, and for the stellar mass, dust mass, and SFR maps. The different panels correspond to the half-light radius $R_{\text{half}}$ normalized by the optical $R_{25}$ diameter (top left), the concentration index (top right), the asymmetry index (middle left), the smoothness index (middle right), the Gini coefficient (bottom left), and the $M_{20}$ indicator (bottom right).}
\label{StatMorph-individual.fig}
\end{figure*}

\begin{figure*}
\centering
\includegraphics[width=\textwidth]{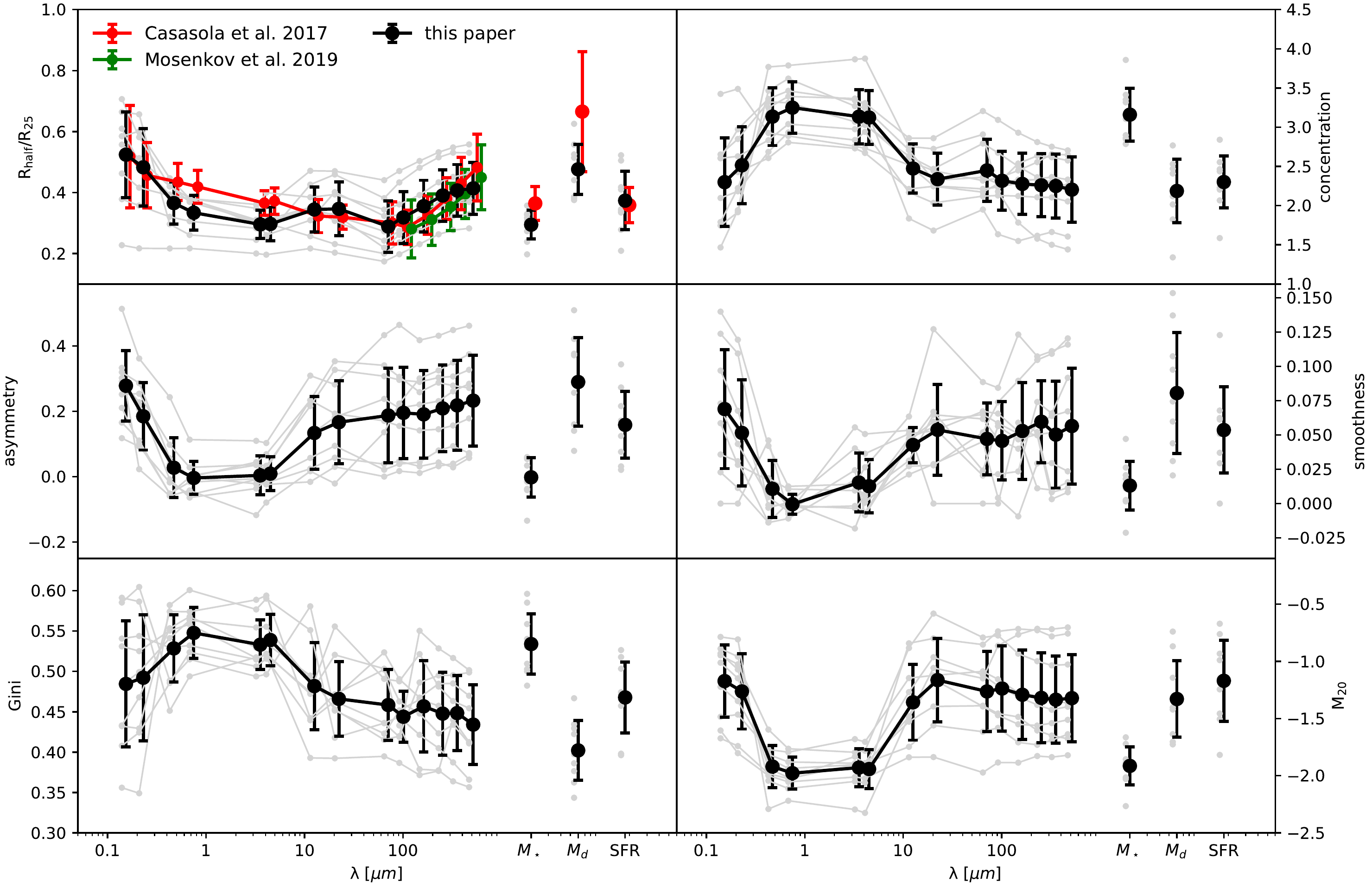}
\caption{Nonparametric morphological indicators for the galaxy sample as a whole as a function of wavelength, and for the stellar mass, dust mass, and SFR maps. The different panels are the same as in Fig.~{\ref{StatMorph-individual.fig}}. The gray lines are the measurements for the individual galaxies in the sample, and the black symbols and their error bars correspond to the mean and the standard deviation of the different indicators over our galaxy sample. }  
\label{StatMorph-joint.fig}
\end{figure*}

We  measured nonparametric morphology indictors for all the maps in a consistent way using the {\tt{StatMorph}} package\footnote{\url{https://statmorph.readthedocs.io/en/latest/}}. This is a user-friendly Python implementation for the calculation of morphological statistics, described in detail by \citet{2019MNRAS.483.4140R}. It is largely based on previous work by \citet{2004AJ....128..163L, 2006ApJ...636..592L, 2008ApJ...672..177L, 2008MNRAS.391.1137L} and has been thoroughly tested and applied in several independent studies \citep[e.g.,][]{2019A&A...632A..98C, 2020MNRAS.491.3624B, 2020MNRAS.494.5636W}.

We applied {\tt{StatMorph}} to the series of maps of each galaxy, with most of the settings fixed at the default values as recommended by \citet{2019MNRAS.483.4140R}. For each galaxy, we created a segmentation map \citep{2004AJ....128..163L} based on the FUV, IRAC 3.6~$\mu$m, and SPIRE 250~$\mu$m images and used it for the analysis of all images and maps of that galaxy.

In our analysis we  concentrate on the following indices:
\begin{description}
\item[\em{Half-light radius}] The half-light radius $R_{\text{half}}$ is calculated as the elliptical radius of the isophote that contains half of the emission. It is probably the most widely adopted indicator for the characteristic size of a galaxy.
\item[\em Concentration] The concentration index is defined as $5\log(R_{80}/R_{20})$, where $R_{20}$ and $R_{80}$ are the elliptical radii of the isophotes that contain 20\% and 80\% of the light, respectively. The concentration of the light distribution in galaxies correlates reasonably well with Hubble type \citep{1958PASP...70..364M, 1959PASP...71..394M, 1984ApJ...280....7O, 2000AJ....119.2645B} and with several other physical properties such as bulge-to-disk ratio, Mg/Fe abundance ratio, and supermassive black hole mass \citep{2001ApJ...563L..11G, 2003ApJS..147....1C, 2004ApJ...601L..33V, 2018MNRAS.477.2399A}.
\item[\em Asymmetry] The asymmetry index is obtained by subtracting the galaxy image rotated by 180$^\circ$ from the original image. This indicator can be used to search for indications of galaxy interactions and mergers, which often reveal strong asymmetry. For normal star-forming galaxies, the asymmetry parameter at optical wavelengths correlates with optical broadband color \citep{1995ApJ...451L...1S, 2000ApJ...529..886C, 2003ApJS..147....1C}. 
\item[\em Smoothness] The smoothness index, sometimes more appropriately called the clumpiness index, was originally introduced as a way to quantify the morphological dichotomy between early-type galaxies without significant ongoing star formation, and late-type galaxies dominated by it. It is obtained by subtracting a smoothed version of the galaxy image from the original image, so galaxies with a larger value for the smoothness index have a less smooth appearance. Smoothness indices based on optical images tend to correlate weakly with optical color and SFR \citep{2003ApJS..147....1C}. 
\item[\em Gini] The Gini coefficient was originally introduced in economics to measure wealth inequality \citep{Gini1912}. In the context of galaxy morphology, it indicates the spread in intensity between the pixels in the aperture. It was introduced by \citet{2003ApJ...588..218A} as a generalized measure of concentration that is applicable to galaxies of arbitrary shape.
\item[\em $M_{20}$ index] The $M_{20}$ indicator measures the second moment of the brightest regions of a galaxy relative to the total second-order central moment. This indicator is also somewhat similar to the concentration index, but it is more sensitive to bright structures outside the center of the galaxy, and therefore relatively sensitive to spiral arms, rings, multiple nuclei, and tidal structures \citep{2004AJ....128..163L, 2014ApJ...781...12H}. The Gini and $M_{20}$ are often used together in order to separate galaxies in different subclasses \citep[e.g.,][]{2008ApJ...672..177L, 2019MNRAS.483.4140R}. 
\end{description}
The {\tt{StatMorph}} package can also be used for S\'ersic profile fitting and to calculate other nonparametric morphological indicators \citep{2007ApJ...656....1L, 2013MNRAS.434..282F, 2016MNRAS.456.3032P}. The six commonly used indices discussed above are sufficient for our purposes.

\section{Results}
\label{Results.sec}

In Fig.~{\ref{StatMorph-individual.fig}} we show the six nonparametric morphology indicators as a function of wavelength, and for the stellar mass, dust mass, and SFR maps, for each of the nine galaxies in our sample. In Fig.~{\ref{StatMorph-joint.fig}} we show the same panel, but we focus on the global trends, which are represented by the black symbols with error bars, representing the mean and standard deviation within the sample.

The half-light radius $R_{\text{half}}$, normalized to the optical $R_{25}$ radius, has a characteristic behavior as a function of wavelength, with large values in the FUV and a gradual decrease over the optical regime to the NIR. In the FIR and submm, $R_{\text{half}}$ increases systematically for all galaxies in our sample, and it reaches   values at 500 $\mu$m similar to those in the FUV band. Not surprisingly, the $R_{\text{half}}/R_{25}$ value for the stellar mass map is similar to the IRAC band values. Interestingly, the half-mass radius of the dust surface density map is larger than the half-light radius measured in any of the individual FIR or submm bands. 

\begin{figure*}
\centering
\includegraphics[width=0.8\textwidth]{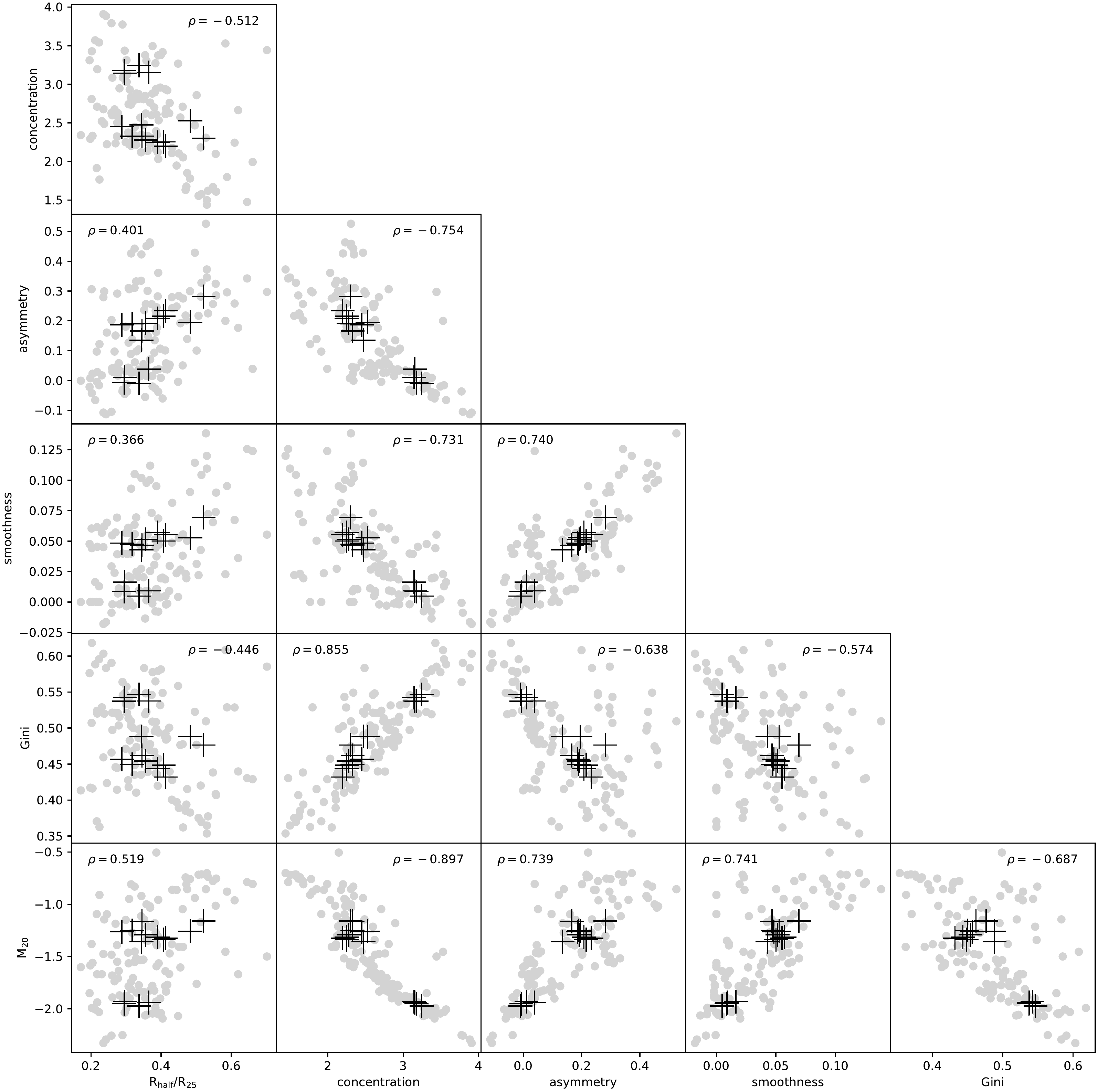}
\hspace*{2em}
\caption{Correlations between the different nonparametric morphology indicators considered in this study. The gray dots correspond to the indicator measurements for the individual galaxies in the different bands, the black crosses are the mean values. Spearman rank coefficients, based on the indicator values of the individual galxy images, are indicated in each panel. }  
\label{StatMorph-corner.fig}
\end{figure*}

The dependence of galaxy size, defined in different ways, as a function of wavelength has been investigated by many authors, and our results are in qualitative agreement with these studies. The steady decline in the galaxy size from the UV to the NIR is due to a stellar mass-to-light ratio gradient, resulting from a combination of intrinsic stellar population gradients and stronger reddening by dust in the inner regions \citep{1996A&A...313..377D, 2003ApJ...582..689M, 2009ApJ...703.1569M, 2011ApJ...731...10M}. Conversely, the steady increase of $R_{\text{half}}/R_{25}$ from mid-IR (MIR) to submm wavelengths is due to a dust temperature gradient, with colder dust residing at larger galactocentric radii \citep{2015A&A...576A..33H, 2016MNRAS.462..331S}. The larger value of $R_{\text{half}}$ from the dust mass map compared to the individual {\it{Herschel}} bands can be understood in the same way. The dust SED fits do indeed reveal a clear temperature gradient for the galaxies in our sample, as discussed by \citet{2017A&A...605A..18C}.

The red circles in the top left panel of Fig.~{\ref{StatMorph-joint.fig}} correspond to results by \citet{2017A&A...605A..18C}. These authors measured the scale-length $h$ as a function of wavelength of their galaxies by fitting an exponential disk to the radial surface brightness profiles. We  converted the scale-lengths for the galaxies in our sample to half-light radii using the relation $R_{\text{half}} = 1.678\,h$ appropriate for exponential disks. The agreement is excellent, except in the optical--NIR region, where \citet{2017A&A...605A..18C} find larger values then we do here. The reason for this discrepancy is that they determine the scale-length of the disk only, whereas we  determined $R_{\text{half}}$ based on the entire light distribution. The presence of a substantial bulge contribution at optical--NIR wavelengths naturally leads to smaller values for $R_{\text{half}}$ in our case. Similarly, the green symbols in the top left panel of Fig.~{\ref{StatMorph-joint.fig}} correspond to the mean effective radii for the galaxies in our sample as determined from S\'ersic fits to the 2D surface brightness maps in five {\it{Herschel}} bands by \citet{2019A&A...622A.132M}. Again, the agreement is excellent, especially taking into account the different methods. 

The concentration index shows a clear pattern when plotted as a function of wavelength for the galaxies in our sample: it generally increases from UV to NIR wavelengths, and subsequently drops back at MIR wavelengths (top right panel of Fig.~{\ref{StatMorph-joint.fig}}). This behavior is in agreement with the results of \citet{2009ApJ...703.1569M}. From the MIR to submm wavelengths, the concentration index is characterized by a relatively modest decrease, due to the decrease in dust temperature with increasing galactocentric radius. In general, the average values in the submm are similar to the FUV value. In agreement with these trends, we find a high concentration in the stellar mass map, and lower values in the dust mass and SFR maps.

The center-left panel of Fig.~{\ref{StatMorph-joint.fig}} shows a very consistent wavelength dependence for the asymmetry index: galaxies are highly asymmetric in the FUV, very symmetric in the NIR,  and the asymmetry then  keeps increasing towards submm wavelengths. This behavior at FUV--MIR wavelengths is in agreement with the results of \citet{2009ApJ...703.1569M}. However, they find a decrease in the asymmetry towards the FIR for most galaxies, in disagreement with our continued increase until the last submm data point. This difference is most likely due to the poor resolution of their {\it{Spitzer}} MIPS data, whereas each galaxy in our limited sample is well resolved even in the SPIRE 500 $\mu$m band. We also note that the asymmetries measured from the dust mass maps are always higher than the asymmetries in each of the individual {\it{Herschel}} images. 

On average, the smoothness index decreases sharply from UV to NIR wavelengths for the galaxies in our sample, implying that galaxies are more clumpy at shorter wavelengths. The smoothness index slightly increases from the NIR all the way to the submm. There is  some diversity in our sample in the smoothness indices at FIR and submm wavelengths. Interestingly, the dust mass map has, again, a higher smoothness index than any of the individual {\it{Herschel}} maps. 

A large spread in Gini indices at UV wavelengths is seen for the galaxies in our sample. On average, the Gini index increases with wavelength until the NIR, where the spread is largely reduced, and subsequently slowly decreases towards 500 $\mu$m. The dust mass maps have, on average, a lower Gini index than any broadband image. Our Gini indices are smaller than those measured by \citet{2007MNRAS.380.1313B}, \citet{2009ApJ...703.1569M}, or \citet{2011MNRAS.416.2401H}. This is most probably related to the pixel size of the images and the aperture within which the index is measured \citep{2011MNRAS.416.2401H, 2019MNRAS.483.4140R}. In our analysis we  used the same pixel size and segmentation map for all wavelengths, such that the measurements at different wavelengths can be directly  compared. 

Finally, for our sample of galaxies, we see a very steady pattern for $M_{20}$ as a function of wavelength. It starts at a high value in the UV, rapidly drops to a minimum at optical--NIR wavelengths, and subsequently increases again in the MIR. This behavior is in agreement with \citet{2009ApJ...703.1569M}. Across the entire MIR--submm range, the $M_{20}$ index is more or less constant, and this is also reflected in the $M_{20}$ indices of the dust mass and SFR maps.  

Comparing the different panels of Figs.~{\ref{StatMorph-joint.fig}}, it is evident that several of the nonparametric morphology indicators have a similar wavelength dependence. This is not surprising, as it has been demonstrated that different indicators are correlated to different degrees \citep{2003ApJS..147....1C, 2004AJ....128..163L, 2007ApJS..172..406S, 2009ApJ...703.1569M}. In Fig.~{\ref{StatMorph-corner.fig}} we show the correlation between the different indicators considered in this study, together with the Spearman rank coefficients. The strongest correlations are those between concentration index on the one hand, and either Gini or $M_{20}$ on the other. The weakest correlations are those between the half-light radius and any of the other nonparametric indicators.

\section{Discussion and conclusions}
\label{Discussion.sec}

This study of a small sample of nearby DustPedia spiral galaxies serves as a proof-of-concept study for the application of nonparametric morphology over the entire UV--submm wavelength range. In particular, it is the first application of these indices to sufficiently resolved FIR and submm maps of galaxies and to the stellar mass, dust mass, and SFR maps of galaxies. 

Nonparametric morphological studies of galaxies at optical wavelengths are usually performed to classify galaxies into different classes and/or to search for signatures of recent interactions \citep{2006ApJ...636..592L, 2007ApJS..172..406S, 2009MNRAS.394.1956C, 2009A&A...497..743H, 2012MNRAS.420..926F, 2014ApJ...781...12H, 2014ARA&A..52..291C}. While galaxy classification could in principle be extended to include morphological information at FIR--submm wavelengths, this is not obvious at all, and this is not the main objective of this study. An in-depth analysis of the nonparametric morphology of the entire DustPedia galaxy sample and the link between morphology and physical properties will be considered in future work.

Our results can have some repercussions on our understanding of the structure of galaxies. Several previous studies have already indicated that star formation is distributed in a more clumpy and less concentrated way compared to stars in galaxies \citep[e.g.,][]{2007MNRAS.380.1313B, 2009ApJ...703.1569M, 2016ApJ...828...27N, 2018MNRAS.477.3014B, 2019A&A...626A..61M}. In this paper we extend these studies to the cool interstellar dust traced by FIR--submm emission. One important lesson that can be learnt from Fig.~{\ref{StatMorph-joint.fig}} is that the interstellar dust in galaxies is distributed in a more extended, less concentrated, more asymmetric, and more clumpy way than the stars. This is clearly visible when comparing morphological indicators at, for example, 3.6 and 250 $\mu$m, or even more outspoken when comparing the indicators corresponding to the stellar and dust mass maps. This should serve as a warning sign against treating the dust in galaxies as a simple smooth component. Even more than stars, the dust distribution differs from a smooth axisymmetric, exponentially declining disk, although for the sake of simplicity this is often a convenient first-order approach \citep[e.g.,][]{1999A&A...344..868X, 2000A&A...359...65B, 2006A&A...456..941M, 2010A&A...518L..39B, 2010MNRAS.403.2053G, 2011A&A...527A.109P, 2017MNRAS.470.2539P, 2014MNRAS.441..869D, 2018A&A...616A.120M}. Detailed modeling of the interaction between dust and starlight on galactic scales should attempt to take into account the more complex morphology of the dust distribution \citep{2014A&A...571A..69D, 2017A&A...599A..64V,  2020arXiv200501720V, 2019MNRAS.487.2753W, 2020A&A...637A..24V, 2020A&A...637A..25N}.

This study also has implications for our understanding of the way galaxies evolve through cosmic times, in particular for numerical simulations of galaxy formation and evolution. Major cosmological hydrodynamics simulations are usually calibrated and tested using statistical distributions and scaling relations based on integrated luminosities or physical properties \citep[e.g.,][]{2014MNRAS.444.1518V, 2015MNRAS.446..521S, 2019MNRAS.486.2827D, 2019MNRAS.490.3196P}. Galaxy morphology, which is the end product of external and internal physical processes alike, can serve as a new and critical test-bed for such studies. Nonparametric morphological indicators have been applied to synthetic images of simulated galaxies to test the fidelity of galaxy formation models \citep{2015MNRAS.451.4290S, 2015MNRAS.454.1886S, 2017MNRAS.465.1106B, 2020MNRAS.491.3624B, 2019MNRAS.483.4140R}. These studies show the power of morphology as a test for galaxy evolution models. In particular,  \citet{2019MNRAS.483.4140R} demonstrate the improvement of the TNG100 simulation \citep{2018MNRAS.475..624N} compared to its predecessor Illustris \citep{2014MNRAS.444.1518V} in terms of the position of galaxies in the Gini--$M_{20}$ plane.

Our study now allows  these morphology tests to be extended for simulated galaxies to wavelength bands beyond the traditional ones dominated by stellar emission. This is particularly timely because of recent developments in cosmological hydrodynamics simulations. The spatial resolution of cosmological hydrodynamics simulations has been steadily increasing over the past few years and, in particular, several suites of high-resolution zoom simulations have been presented \citep{2014MNRAS.445..581H, 2015MNRAS.454...83W, 2017MNRAS.467..179G, 2020arXiv200401914F}. A complementary development is the incorporation of dust physics into hydrodynamics simulations \citep{2017MNRAS.468.1505M, 2018MNRAS.478.2851M, 2019MNRAS.486.2827D, 2020MNRAS.491.3844A}. In combination with the improved accuracy and power of radiative transfer post-processing algorithms \citep{2010MNRAS.403...17J, 2016MNRAS.462.1057C, 2018ApJS..234...20C, 2017MNRAS.470..771T}, this makes the morphology of galaxies from UV to submm wavelengths an interesting test for cosmological hydrodynamics simulations. 

\bibliography{StatMorph}

\end{document}